\documentclass{article}
\usepackage{amsfonts}

\input tcilatex

\begin{document}

\title{Maps and fields with compressible density}
\author{Thomas H. Otway\thanks{%
email: otway@ymail.yu.edu} \\
\\
\textit{Department of Mathematics, Yeshiva University,}\\
\textit{\ New York, New York 10033 U.S.A.}}
\date{}
\maketitle

\begin{abstract}
Properties of steady compressible flow for which geometric constraints have
been placed on the potential function are derived, under hypotheses on the
flow density and the singular set. Some related unconstrained problems are
also considered, including the estimation of a class of fields having
nonzero vorticity. 2000 \textit{MSC: 58E20, 58E99, 75N10.}
\end{abstract}

\section{Introduction}

The study of certain classical fields leads to a generalization of harmonic
maps in which the Dirichlet energy is replaced by the functional 
\[
E=\int_{M}\int_{0}^{Q(du)}\rho \left( s\right) dsdM; 
\]
here $M$ is a Riemannian manifold; $du$ is the differential of a map $%
u:M\rightarrow N,$ where $N$ is another Riemannian manifold; 
\[
Q\left( du\right) =\left\langle du,du\right\rangle _{\mid T^{\ast }M\otimes
u^{-1}TN}; 
\]
$\rho :\Bbb{R}^{+}\cup \left\{ 0\right\} \rightarrow \Bbb{R}^{+}$ is a $%
C^{1,\alpha }$ function of $Q\ $satisfying the differential inequality 
\begin{equation}
0<\frac{\frac{d}{dQ}\left[ Q\rho ^{2}(Q)\right] }{\rho \left( Q\right) }%
<\infty
\end{equation}
for $Q\in \lbrack 0,Q_{crit}).$

In the typical case, the manifold $N$ represents a geometric constraint
placed on the flow potential of a steady, irrotational, polytropic ideal
fluid for which the closed 1-form $du\in T^{\ast }M$ is dual to the flow
velocity. \ In this case we choose 
\begin{equation}
\rho \left( Q\right) =\left( 1-\frac{\gamma -1}{2}Q\right) ^{1/\left( \gamma
-1\right) },
\end{equation}
where $\gamma >1$ is the adiabatic constant of the medium and $Q_{crit}$ is
the square of the sonic flow speed. These choices transform (1) into a
condition for subsonic flow of mass density $\rho .$

The study of functionals of this kind in the unconstrained Euclidean case,
in which $N=\Bbb{R}^{k}$ and $M$ is a domain of $\Bbb{R}^{n},$ goes back at
least to work on planar flow by Bateman in the 1920s [\textbf{Ba}]. An
extensive bibliography covering the first half of the last century is given
in [\textbf{Be}]. A more recent bibliography of mathematical work in
compressible fluid dynamics (not necessarily connected with variational
theory) appears in [\textbf{Ch}]; see also [\textbf{DO}] and the
bibliographic remarks in [\textbf{CF}].

A discussion of unconstrained compressible flow in a local chart on a
manifold appears in Sec. III.3 of [\textbf{Sed}]. A global existence theorem
for steady, unconstrained subsonic flow on a compact Riemannian manifold is
given in [\textbf{SS1}]; for subsequent research employing this \textit{%
nonlinear Hodge} approach see, \textit{e.g.}, [\textbf{SS2}], [\textbf{SS3}%
], [\textbf{Si}], [\textbf{Sm}], and [\textbf{ISS}]. In those works the
curvature of the manifold introduces geometry into the domain of the
velocity field. By considering potentials subject to a geometric constraint,
as we do here and in [\textbf{O1}], we introduce geometry into the range of
the velocity field. In [\textbf{O1}] we emphasized this connection to the
preceding literature by calling such potentials \textit{nonlinear Hodge maps.%
} But the potentials studied in [\textbf{O1}] are not associated with a
cohomology class and neither the geometric construction nor the physical
interpretation extend automatically to higher-degree forms. So it is perhaps
more accurate to call mappings which are critical points of $E$ \textit{%
compressible-density maps.}

There is already a considerable literature on maps for which a nonquadratic
energy functional is given by the $L^{p}$ norm of the gradient of the map $%
u; $ see, \textit{e.g.}, [\textbf{FH}], [\textbf{HL}], [\textbf{F}] and the
references therein. \ Those works are motivated by the mathematical
observation that the harmonic map energy is the squared $L^{2}$ norm of the
gradient, raising the question of whether a corresponding theory can be
derived for stationary points of the nonmetrizable $L^{p}$ norms, $p>1.$ \
Our starting point, on the other hand, is the physical observation that
harmonic maps model a geometric constraint on a field of constant mass
density. \ This prompts one to ask whether a corresponding theory can be
derived for fields having mass density $\rho $ which depends in a nonlinear
way on field strength. That question leads to the replacement of the
harmonic map energy by the functional $E.$ In applications to fluid
dynamics, the variational equations of $E$ correspond to continuity
equations for the flow. In the incompressible (hydrodynamic) limit $\rho
\equiv 1,$ the variational equations reduce to the harmonic map equations.
This incompressible special case has also been interpreted in the context of
nonlinear elasticity [\textbf{T}].

The analysis of critical points of $E$ is necessarily somewhat different
from analysis of the harmonic map energy or of the other $L^{p}$ norms of
the gradient. In those cases scaling arguments are natural, whereas they may
be unnatural for $E,$ as they involve a choice of conformal behavior for $%
\rho .$ This is a point of commonality between our problem and certain other
recent extensions of the harmonic map equations, \textit{e.g.,} [\textbf{A}]
and [\textbf{LM}]; see also [\textbf{EL}]. The density of maps which are $%
L^{p}$-critical points of their gradient tends to zero \textit{(cavitates)}
as ellipticity degenerates; this behavior is atypical of the mass density of
fluids, for which the sonic value lies at the supremum of the range of
subsonic speeds. [Compare conditions (2), above, and (23), below]. Finally,
the references cited at the beginning of the preceding paragraph assume an
energy minimizing property. We assume only that Euler-Lagrange equations are
satisfied on a given subdomain.

In Section 2 an $L^{\infty }$ estimate is derived for nonuniformly elliptic,
scalar velocity fields. Section 3 concerns technical aspects of the
uniformly elliptic case which we studied earlier ([\textbf{O1}], Theorem 3).
We also present in Section 3 a somewhat different proof of the result in [%
\textbf{O1}], one which is simpler in that it avoids certain smoothness
assertions which were necessary in the original argument. In Section 4,
Corollary 8 of [\textbf{O1}] is extended from the uniformly elliptic case to
the nonuniformly elliptic case under somewhat different hypotheses.

We note that the analysis literature tends to treat the velocity as a
section of the cotangent bundle, whereas the physics and geometry literature
puts this object in the tangent bundle. The local arguments of Sec. 3 are
the same in both representations. The usefulness of the cotangent
representation for considering fields with vorticity is apparent in Sec. 4;
for consistency we employ this notation in Sec. 3 as well.

\section{Near-sonic maps into a line}

It is known [\textbf{D}] that if $u\in H^{1,p}(\Omega )$ is a weak solution
of the scalar equation 
\begin{equation}
div\left( \left| \nabla u\right| ^{p-2}\nabla u\right) =0
\end{equation}
for $p>1$ in an open domain $\Omega $ of $\Bbb{R}^{n}$, then for every $n$%
-disc $B_{R}\subset \Omega $ of radius $R$ and every number $\delta \in
\left( 0,1\right) $ there is a constant $k(\delta )$ independent of $R$ for
which 
\begin{equation}
\left\| \nabla u\right\| _{L^{\infty }\left( B_{R-\delta R}\right)
}^{p/2}\leq k(\delta )R^{-N}\int_{B_{R}}\left| \nabla u\right| ^{p}dx.
\end{equation}
(See also [\textbf{E}], [\textbf{Le}], and references therein.)

This is a useful result to have, as the semi-norm on the right is the energy
integral associated to weak solutions of (3). \ Thus inequality (4) derives
an $L^{\infty }$ bound on the weak solution, which is unnatural to impose
directly, from a condition of finite energy, which is the natural condition
to impose on solutions of equations with variational structure. A uniform $%
L^{\infty }$ bound on weak solutions plays important roles in\ smoothness
estimates and numerical analysis. \ Inequality (4) can be significantly
generalized within the class of $L^{p}$-stationary gradients (see, \textit{%
e.g.}, [\textbf{HL}]).

In this section we derive an analogue of inequality (4) for solutions of the
scalar equation 
\begin{equation}
div\left[ \rho \left( Q\right) \nabla u\right] =0
\end{equation}
for which $\rho $ satisfies condition (1). Equation (5) is the
Euler-Lagrange equation for the functional E in the special case in which $u$
is a scalar function on $\Bbb{R}^{n}.$

If the middle term of (1) is bounded below away from zero on the entire
range of values for $Q$, then this analogue has already been derived in
considerable generality (see, \textit{e.g.}, Theorem 4.3 of [\textbf{SS}],
Proposition 3.1 of [\textbf{Si}],\ Lemma 3 of [\textbf{DO}], or Theorem 9 of
[\textbf{O1}]). \ In each of these cases, however, the constant analogous to 
$k(\delta )$ of inequality (4) tends to infinity as $Q$ tends to $Q_{crit},$
so these inequalities are not uniform unless eq. (5) is itself uniformly
elliptic. \ Rather, the cited inequalities contribute indirectly to uniform
H\"{o}lder estimates, by way of a delicate limiting argument introduced by
Shiffman [\textbf{Sh}] in the planar case and extended to higher dimensions
in [\textbf{SS}]. \ Direct arguments should suffice to estimate weak
solutions of eq. (5) which, unlike the equations studied in the works cited,
has scalar solutions. \ Our goal is to derive estimates for (5) which are
manifestly uniform over the entire subcritical range of values for $Q.$

By a \textit{weak solution} in this scalar case we mean a function $u$
having finite energy $E$ and satisfying $\forall $ $\psi \in
H_{0}^{1}(\Omega )$ the integral identity 
\begin{equation}
\int_{\Omega }\left\langle \rho (Q)du\,,\,d\psi \right\rangle \,dx=0,
\end{equation}
where $x$ is a vector in a bounded type-A domain $\Omega \subset \Bbb{R}%
^{n}, $ $n>2,$ and the angle brackets denote the euclidean inner product on
1-forms. For a definition of \textit{type-A domain} see, \textit{e.g.,} p.
68 of [\textbf{G}]; our intention is to insure that a ball in the interior
of $\Omega $ does not become trapped in an outward cusp. As an example, any
Lipschitz domain is type-\textit{A}. In order for the following theorem to
make sense in terms of fluid dynamics, we must additionally impose the
condition that $\Omega $ be topologically trivial in order that the flow
potential remain single-valued.

\begin{theorem}
Let the scalar function $u(x),$ $x\in \Omega ,$ be a weak solution in the
sense of (6) for $\rho $ satisfying condition (1). \ In addition, assume
that $\rho ^{\prime }(Q)\leq 0\ \forall Q\in \lbrack 0,Q_{crit})$ and that
on this range, $0<\kappa _{0}\leq \rho (Q)\leq \rho (0)<\infty .$ \ \ Then
for every $n$-disc $B_{R}$ strictly contained in $\Omega $ and every $\delta
\in \left( 0,1\right) $ there exists a constant $\kappa _{1}(n,\delta
,\kappa _{0},\rho (0))$ for which 
\begin{equation}
\sup_{x\in B_{R\left( 1-\delta \right) }}Q(x)\leq \kappa
_{1}R^{-n}E_{|B_{R}}.
\end{equation}
\end{theorem}

The constant $\kappa _{1}$ in Theorem 1 depends neither on the radius $R$
nor on any ellipticity parameter [such as the parameter $\kappa _{3}$ of
condition (15), below]. \ \ Thus in particular, inequality (7) does not
necessarily follow from the uniform bound of $\sqrt{2/\left( \gamma
+1\right) }$ on the subsonic flow speed in (2). At the same time, it is
satisfying to have a bound on weak solutions that results only from
mathematical hypotheses on the equation itself rather than relying on a
bound, such as the sonic speed, which is imposed on the mathematics by a
physical model. The noncavitation hypothesis and the other hypotheses of the
theorem are satisfied on the subsonic range by mass densities of the form
(2).

The proof of Theorem 1 is elementary. \ The idea is to choose the test
function in (6) to be a local restriction of the antiderivative $F(Q)$ for
the function 
\[
f(Q)\equiv \rho ^{2}(Q)+2Q\rho (Q)\rho ^{\prime }(Q). 
\]
The ellipticity condition is then interpreted, where it appears, as a piece
of the chain rule applied to the gradient of $F$. \ This relieves us of the
necessity to bound $f(Q)$ below away from zero, but obliges us to translate
statements about $F(Q)$ and its $L^{2}$-norm into statements about $Q$ and
its energy functional. \ Such an approach combines ideas from Sec. 1 of [%
\textbf{U}] and Sec. 3 of [\textbf{D}]. Those papers, as well as [\textbf{Ev}%
] and [\textbf{Le}], concern weak solutions satisfying hypotheses similar to
inequality (26) of Sec. 4.

\bigskip

\textit{Proof}. We initially assume that $u$ is twice continuously
differentiable. \ It is then easy to verify that the results are unaffected
if the derivatives are replaced by limits of finite differences. \ Taking a
weak derivative of (6) yields

\begin{equation}
\sum_{i=1}^{n}\int_{\Omega }\left( \rho (Q)u_{x^{i}}\right) _{x^{j}}\psi
_{x^{j}}\,dx=0
\end{equation}
where, here and below, repeated indices are summed from 1 to $n$. \ For a
function $\zeta \in C_{0}^{\infty }\left( B_{R}\right) $ and positive
parameters $\alpha $ and $\beta ,$ choose 
\[
\psi (x)=u_{x^{i}}\left[ Q\rho ^{2}(Q)+\beta \right] ^{\alpha /2}\zeta
^{2}(x). 
\]
Expanding the integrand of (8) yields a sum of six terms: 
\[
\sum_{i=1}^{n}\left( \rho (Q)u_{x^{i}}\right) _{x^{j}}\psi _{x^{j}}=\rho
^{\prime }(Q)Q_{x^{j}}u_{x^{i}}u_{x^{i}x^{j}}\left[ Q\rho ^{2}(Q)+\beta %
\right] ^{\alpha /2}\zeta ^{2}+ 
\]
\[
\frac{\alpha }{2}\rho ^{\prime }(Q)Q_{x^{j}}u_{x^{i}}u_{x^{i}}\left[ Q\rho
^{2}(Q)+\beta \right] ^{\left( \alpha -2\right) /2}\left[ \rho
^{2}(Q)+2Q\rho (Q)\rho ^{\prime }(Q)\right] Q_{x^{j}}\zeta ^{2} 
\]
\[
+\rho ^{\prime }(Q)Q_{x^{j}}u_{x^{i}}u_{x^{i}}\left[ Q\rho ^{2}(Q)+\beta %
\right] ^{\alpha /2}2\zeta \zeta _{x^{j}} 
\]
\[
+\rho (Q)u_{x^{i}x^{j}}u_{x^{i}x^{j}}\left[ Q\rho ^{2}(Q)+\beta \right]
^{\alpha /2}\zeta ^{2}+ 
\]
\[
\frac{\alpha }{2}\rho (Q)u_{x^{i}x^{j}}u_{x^{i}}\left[ Q\rho ^{2}(Q)+\beta %
\right] ^{\left( \alpha -2\right) /2}\left[ \rho ^{2}(Q)+2Q\rho (Q)\rho
^{\prime }(Q)\right] Q_{x^{j}}\zeta ^{2} 
\]
\[
+\rho (Q)u_{x^{i}x^{j}}u_{x^{i}}\left[ Q\rho ^{2}(Q)+\beta \right] ^{\alpha
/2}2\zeta \zeta _{x^{j}}\equiv \sum_{r=1}^{6}i_{r}. 
\]
We estimate the terms of this sum individually. \ The following estimates
should be interpreted as occurring ``under the integral sign.'' \ The
hypothesis on the sign of $\rho ^{\prime }(Q)$ implies, using Kato's
inequality, that 
\[
i_{1}+i_{4}\geq \frac{\kappa _{0}}{4\rho ^{2}\left( 0\right) }\left( \frac{4%
}{\alpha +2}\right) ^{2}\left| \nabla \left( \left[ Q\rho ^{2}(Q)+\beta %
\right] ^{\left( \alpha +2\right) /4}\right) \right| ^{2}\zeta ^{2}. 
\]
Because the range of $\rho (Q)/\rho (0)$ is contained in the interval $%
(0,1], $ we also have 
\[
i_{2}+i_{5}\geq 
\]
\[
\frac{\alpha }{4\rho (0)}\left( \frac{4}{\alpha +2}\right) ^{2}\left| \nabla
\left( \left[ Q\rho ^{2}(Q)+\beta \right] ^{\left( \alpha +2\right)
/4}\right) \right| ^{2}\zeta ^{2}. 
\]
Moreover, there exists a positive constant $\varepsilon $ for which 
\[
i_{3}+i_{6}\geq 
\]
\[
-\left( \varepsilon \rho ^{2}/2\right) \left[ \rho (Q)+2Q\rho ^{\prime }(Q)%
\right] ^{2}\left| \nabla Q\right| ^{2}\left[ Q\rho ^{2}(Q)+\beta \right]
^{\left( \alpha -2\right) /2}\zeta ^{2} 
\]
\[
-2(\varepsilon \rho ^{2})^{-1}\left| \nabla \zeta \right| ^{2}\left[ Q\rho
^{2}(Q)+\beta \right] ^{\left( \alpha +2\right) /2}\geq 
\]
\[
-\frac{\varepsilon }{2}\left( \frac{4}{\alpha +2}\right) ^{2}\left| \nabla
\left( \left[ Q\rho ^{2}(Q)+\beta \right] ^{\left( \alpha +2\right)
/4}\right) \right| ^{2}\zeta ^{2} 
\]
\[
-2(\varepsilon \kappa _{0}^{2})^{-1}\left| \nabla \zeta \right| ^{2}\left[
Q\rho ^{2}(Q)+\beta \right] ^{\left( \alpha +2\right) /2}. 
\]
Choose $\varepsilon $ to equal $\left[ \kappa _{0}+\alpha \rho (0)\right]
/4\rho ^{2}(0).$ \ Then we obtain the integral inequality 
\begin{equation}
\int_{\Omega }\left| \nabla \left( \left[ Q\rho ^{2}(Q)+\beta \right]
^{\left( \alpha +2\right) /4}\right) \right| ^{2}\zeta ^{2}\ast 1\leq
m\int_{\Omega }\left( \left[ Q\rho ^{2}(Q)+\beta \right] ^{\left( \alpha
+2\right) /4}\right) ^{2}\left| \nabla \zeta \right| ^{2}\ast 1
\end{equation}
for 
\[
m=\left[ \frac{2\rho ^{2}\left( 0\right) }{\kappa _{0}}\left( \frac{\alpha +2%
}{\kappa _{0}+\alpha \rho (0)}\right) \right] ^{2}. 
\]
As $\alpha $ tends to either zero or infinity, $m$ tends to a finite
constant $\kappa _{2}$ which depends only on the upper and lower bounds on $%
\rho (Q).$

Apply inequality (9.5.8) of [\textbf{LU}] to expression (9), taking the
quantities $u$ and $\varepsilon $ of that reference to equal, respectively,
the quantities $\sqrt{Q\rho ^{2}(Q)+\beta }$ and $\alpha /2$ of expression
(9). \ Construct a Moser iteration along the lines of expressions
(9.5.8)-(9.5.12) in [\textbf{LU}]. We obtain in the limit the inequality 
\begin{equation}
\sup_{x\in B_{R\left( 1-\delta \right) }}\left[ Q(x)\rho ^{2}(Q(x))+\beta %
\right] \leq \kappa _{2}R^{-n}\int_{B_{R}}\left[ Q\rho ^{2}(Q)+\beta \right]
\ast 1.
\end{equation}
We have 
\[
\int_{B_{R}}Q\rho ^{2}(Q)\ast 1=\int_{B_{R}}\int_{0}^{Q}\frac{d}{ds}\left(
s\rho ^{2}(s)\right) \,ds\ast 1\leq 
\]
\begin{equation}
\int_{B_{R}}\int_{0}^{Q}\rho ^{2}(s)\,ds\ast 1\leq \int_{B_{R}}\rho
(0)\int_{0}^{Q}\rho (s)\,ds\ast 1=2\rho (0)E_{|B_{R}}.
\end{equation}
Regarding the left-hand side of inequality (10), condition (1) implies that $%
Q\rho ^{2}(Q)$ is an increasing function of $Q$. \ Thus the \textit{suprema}
in $B_{R\left( 1-\delta \right) }$ of $Q(x)$ and of $Q\rho ^{2}(Q(x))$ occur
at the same value of $x$. \ Because the mass density is noncavitating, 
\begin{equation}
\kappa _{0}^{2}\sup_{x\in B_{R\left( 1-\delta \right) }}Q(x)\leq \sup_{x\in
B_{R\left( 1-\delta \right) }}\left[ Q(x)\rho ^{2}(Q(x))\right] .
\end{equation}
Comparing inequalities (10)-(12), we conclude that there is a constant $%
\kappa _{1}$ such that 
\[
\sup_{x\in B_{R\left( 1-\delta \right) }}Q(x)+\beta \leq \kappa _{1}R^{-n} 
\left[ E_{|B_{R}}+\beta vol(B_{R})\right] , 
\]
where $\kappa _{1}$ depends on $n,\delta ,\kappa _{0},$ and $\rho (0).$ \
Because $\beta $ is an arbitrary positive number, we can let it tend to zero
without affecting the other constants.

We now remove the smoothness assumption on $u$. \ Replace the admissible
test function $\psi (x)$ in eq.\ (6) by the admissible test function $\psi
\left( x-he_{j}\right) ,$ where $e_{j}$ is the $j^{th}$ basis vector for $%
\Bbb{R}^{n},$ $j=1,...,n$, and $h$ is a positive constant. \ Then (6)
assumes the form 
\begin{equation}
\int_{B}\left\langle \rho (Q(x))du(x)\,,\,d\psi \left( x-he_{j}\right)
\right\rangle \,dx=0.
\end{equation}
Apply the coordinate transformation $y=x-he_{j}$ to eq. (13) and\ subtract
(6) from (13) to obtain 
\[
h^{-1}\int_{B}\left\langle \rho (Q(x+he_{j}))du(x+he_{j})-\rho
(Q(x))du(x)\,,\,d\psi \left( x\right) \right\rangle \,dx=0. 
\]
The limiting case is an expression of the form (8). \ The expressions
leading to inequality (9) remain true in the finite difference
approximation. \ Because the right-hand side of (9) does not depend on $h,$
we can allow the parameter $h$ to tend to zero in this approximation,
completing the proof of Theorem 1.

\bigskip

\textbf{Remarks}. \ 1. \ In Sec. 9.5 of [\textbf{LU}] the Moser iteration is
illustrated for linear equations of the form 
\[
\left( \rho _{ij}(x)u_{x^{j}}\right) _{x^{i}}=0, 
\]
where 
\[
\nu \sum_{i=1}^{n}\xi _{i}^{2}\leq \rho _{ij}(x)\xi _{i}\xi _{j}\leq \mu
\sum_{i=1}^{n}\xi _{i}^{2}. 
\]
In this case noncavitation is equivalent to ellipticity, whereas the two
conditions are distinct for the quasilinear density $\rho (Q).$ \ Thus the
ratio $\mu /\nu $ in expression (9.5.8) of [\textbf{LU}], which is analogous
to the factor $m$ in our expression (9), introduces a dependence on
ellipticity in the linear case but not in the quasilinear case.

2. Theorem 9 of [\textbf{O1}] is a subparabolic Moser estimate for
multivalued flow potentials possessing geometric constraints. \ The
preceding proof is too simple to work in that case, and the constants
obtained in the proof of Theorem 9 depend on ellipticity. \ However, one can
replace, in the line preceding inequality (69) of that proof, the function $%
Q^{r-1}$ for $r>2$ by the function $\left( Q+\beta \right) ^{r-1}$ for $r>1$%
, allowing $\beta $ to tend to zero at the end as in the preceding proof. \
This avoids eventual difficulties in the Moser iteration.

\section{Uniformly elliptic maps}

We now consider the more difficult cases in which the target has nontrivial
geometry. In what follows the symbol $C$ will denote generic positive
constants unless otherwise indicated.

\subsection{Effects of geometric constraints}

In studying critical points of $E,$ it is natural to obtain the admissible
class of maps from the condition of finite energy. \ We seek a class of
bounded maps having integrable density $e(u).$ \ In order to integrate this
object, it is necessary to choose local coordinates for $e(u)$ on $N$ and it
is not \textit{a priori} true that $u$ takes a coordinate chart on $M$ into
a coordinate chart on $N$. \ If however we restrict our attention to maps
from $M$ into $N$ which are H\"{o}lder continuous, then the local
oscillations of the map are controlled on the target, and the image of a
sufficiently small region of $M$ will lie in a coordinate chart of $N.$ In
this case we can write 
\[
Q=\frac{1}{2}\gamma ^{\alpha \beta }(x)g_{ij}\left( u(x)\right) \frac{%
\partial u^{i}}{\partial x^{\alpha }}\frac{\partial u^{j}}{\partial x^{\beta
}}, 
\]
where for $n=dim(M),$ $x=(x^{1},...,x^{n})$ is a coordinate chart on the
manifold$\ M$ having metric tensor $\gamma _{\alpha \beta }(x);$ $%
u=(u^{1},...,u^{m})$ is a coordinate chart on the manifold$\ N$ \ having
metric tensor $g_{ij}(u),$ where $m=dim(N);$ repeated Greek indices are
summed from 1 to $n$; repeated Latin indices are summed from 1 to $m$.

This continuity assumption severely restricts the kinds of questions that we
can ask about the map. Moreover, the geometric constraint re-emerges as a
problem, even if the map is continuous, when we attempt to extremize the
energy functional by taking variations. \ This is because the test functions 
$\psi $ might take the image of $u+t\psi $ off of $N$, even for small values
of $t.$ This can be immediately seen if, for example, we take $N$ to be the
unit sphere $\left| u\right| =1.$

The conventional solution to both problems, that of defining an admissible
class of finite-energy maps and of varying the energy on the target
manifold, is to embed $N$ isometrically into a higher-dimensional Euclidean
space $\Bbb{R}^{k}$ by the Nash Embedding Theorem. \ The manifold $N$
emerges as a system of $k-m$ independent constraints, 
\[
\Phi \left( u\right) =\left( \Phi _{1}\left( u\right) ,...,\Phi _{k-m}\left(
u\right) \right) =0. 
\]
In this case 
\[
Q=\frac{1}{2}\gamma ^{\alpha \beta }(x)\frac{\partial u^{i}}{\partial
x^{\alpha }}\frac{\partial u^{i}}{\partial x^{\beta }} 
\]
and the incompressible energy integral reduces to the classical Dirichlet
integral.\ In taking variations, a suitable Euclidean neighborhood $\mathcal{%
O}(N)$ of $N$ is projected onto $N$ by nearest point projection $\Pi $. \ If 
$t$ is small enough and $N$ is a $C^{1}$ submanifold of $\Bbb{R}^{k}$, then
the variations $\Pi \circ (u+t\psi )$ will be constrained to lie on $N$ for
almost every $x$ in $M$, where $\psi \in C_{0}^{\infty }\left( M,\Bbb{R}%
^{k}\right) ,$ and for every $x$ in $M$ if $u$ is continuous. Now the
variational equations of $E$ are given by 
\[
\frac{d}{dt}_{|t=0}\int_{M}\int^{Q_{t}}\rho \left( s\right) dsdM=0, 
\]
where 
\[
Q_{t}=\left| d\left[ \Pi \circ (u+t\psi )\right] \right| ^{2}. 
\]
The variational equations in the ambient space assume the explicit form 
\begin{equation}
\delta \left[ \rho \left( Q\right) du\right] =\rho \left( Q\right) A\left(
du,du\right) ,
\end{equation}
where $\delta $ is the formal adjoint of the exterior derivative $d$ and $A$
is the second fundamental form of $N.$

See [\textbf{Sch}] \ and [\textbf{ScU}] for detailed discussions of these
issues in the harmonic map case.

We call $u\in L^{\infty }\left( M,\Bbb{R}^{k}\right) \cap H^{1,2}\left( M,%
\Bbb{R}^{k}\right) $\ a \textit{weak solution} of eqs. (14) in a coordinate
chart $\Omega $ of $M$ if $u$ has finite energy $E$ and satisfies, $\forall $
$\zeta \in H_{0}^{1,2}(\Omega ,\Bbb{R}^{k})\cap L^{\infty }\left( \Omega ,%
\Bbb{R}^{k}\right) ,$ the identity 
\[
\int_{\Omega }\left\langle d\zeta ,\rho (Q)du\right\rangle \ast
1=\int_{\Omega }\left\langle \zeta ,\rho (Q)A\left( du,du\right)
\right\rangle \ast 1. 
\]
The existence of weak solutions to the unconstrained problem in the elliptic
range follows, by lower semicontinuity, from the convexity of the energy
functional under condition (1). \ Weak solutions of the constrained problem
may not exist for certain choices of $\rho $ and $N$. To see this, let $\rho
(Q)=Q^{(p-2)/2},$ $p>1,$ and consider the counterexample of [\textbf{HL}],
Sec. 6.3.

\subsection{Maps with apparent singularities}

The literature on removable singularities is too large for even a
superficial review. We mention that the removability of singularities in
harmonic maps is considered in, \textit{e.g.,} [\textbf{SaU}], [\textbf{EP}%
], [\textbf{Me}], [\textbf{Li}], and [\textbf{CL}]. \ Obstacles to the
extension of methods used in those references to our case include, in
addition to the dependence of the scaling behavior of $E$ on the choice of $%
\rho ,$ the absence of an obvious analogue to the \textit{a priori}
H\"{o}lder estimate of [\textbf{HJW}], which forms the basis for many
smoothness results in the harmonic map literature. Removable singularities
theorems and related \textit{a priori} estimates for mappings which are
critical points for the $L^{p}$-norm of their gradient are reviewed in [%
\textbf{F}]. Those arguments also strongly depend on the scaling behavior of
the energy. Removability of singularities in systems which resemble the
unconstrained case of eqs. (14) can be found in, for example, [\textbf{ISS}%
]. The application of such results to the constrained case is limited by the
presence of quadratic nonlinearities arising from the target curvature.

The removability of an apparent singularity can be proven either by showing
the existence of a continuous transformation to a nonsingular domain, or by
ruling out the existence of the singular set on \textit{a priori} grounds.
We adopt the latter approach in the following theorem.

\begin{theorem}
Let $u:\Omega \rightarrow N$ be a $C^{2}$ stationary point of the energy $E$
on $\Omega /\Sigma ,$ where $\Omega $ is an open bounded, type-A\ domain of $%
\Bbb{R}^{n}$, $\,n>2;$ $N$ is a smooth, compact $m$-dimensional Riemannian
manifold, $m\leq n,$ $\partial N=0;$ $\Sigma \subset \subset B\subset
\subset \Omega $ is a compact singular set, completely contained in a
sufficiently small \textit{n}-disc $B$, which is itself completely contained
in $\Omega .$\ \ Suppose that $\rho $ satisfies 
\begin{equation}
\kappa _{3}<\rho (Q)+2Q\rho ^{\prime }(Q)<\kappa _{4}
\end{equation}
for constants $0<\kappa _{3}<\kappa _{4}<\infty .$ \ If $n>4$, let $%
2n/(n-2)<\mu \leq n,$ where $\mu $ is the codimension of $\Sigma ,$ and let $%
du$ $\in $ $L^{n}(B);$ if $n=3,4,$ let $du\in L^{4q_{0}\beta }(B)\cap
L^{4q}(B),$ where $\beta =\left( \mu -\varepsilon \right) /\left( \mu
-2-\varepsilon \right) $ for $2<\mu \leq n,$ $\varepsilon >0,$ and $\frac{1}{%
2}<q_{0}<q.$ Then $du$ is H\"{o}lder continuous on $B.$
\end{theorem}

Because the singular set is assumed small, the choice of a Euclidean domain $%
\Omega $ entails little reduction in generality. In distinction to the
harmonic map case, Theorem 2 does not immediately imply any higher degree of
smoothness. The theorem immediately extends to the case of a finite number
of singular sets having the same properties as $\Sigma .$

Theorem 2 is stated and proven in [\textbf{O1}] (Theorem 3). We begin by
briefly reviewing that proof, adding details on the underlying elliptic
theory in Lemmas 4 and 5. An alternate method of proof, which avoids Lemmas
4 and 5 altogether, is given in Sec. 3.3.1. We show in the proof that the
modulus of continuity for $du$ depends on $\rho ,u,N,$ $n,$ and on the $%
L^{n} $-norm of $du.$ A metric can be chosen on $\Omega $ in which the $%
L^{n} $-norm of $du$ over $\Omega $ is smaller than any given fixed number.
There are choices of $\rho ,$ however, under which the variational equations
fail to be invariant under this transformation; \textit{c.f.} [\textbf{KFL}].

\begin{lemma}
Under the hypotheses of Theorem 2, $u$ is H\"{o}lder continuous on $B.$
\end{lemma}

\textit{Proof.} Away from the singular set, $Q$ is sufficiently smooth that
local coordinates can be chosen on $N$ and one can show ([\textbf{O1}],
Theorem 2) that 
\begin{equation}
L(Q)+C_{N}Q\left( Q+1\right) \geq 0,
\end{equation}
where $L$ is an elliptic operator under hypothesis (15). Integrate
inequality (16) by parts over $B$ against a test function $\left( \eta \psi
\right) ^{2}\Xi (Q);$ here $\eta ,\psi \geq 0;\psi (x)=0\,\forall x$ in a
neighborhood of $\Sigma ;\eta \in C_{0}^{\infty }(B^{\prime })$ where $%
\Sigma \subset \subset B^{\prime }\subset \subset B;$ $\Xi (Q)=h(Q)h^{\prime
}\left( Q\right) ,$ where for $k=0,1,\ldots ,$ 
\[
h(Q)= 
\]
\[
\left\{ 
\begin{array}{l}
Q^{[n/(n-2)]^{k}n/4}\;for\;0\leq Q\leq \ell , \\ 
\frac{\mu -\varepsilon }{\mu -2-\varepsilon }\left[ \left( \ell \cdot
Q^{(\mu -2-\varepsilon )/2}\right) ^{[n/(n-2)]^{k}n/2(\mu -\varepsilon )}-%
\frac{2}{\mu -\varepsilon }\ell ^{\lbrack n/(n-2)]^{k}n/4}\right] for\;Q\geq
\ell
\end{array}
\right. 
\]
if $n>4;$ $h$ is an analogous test function ([\textbf{Se}], p. 280) when $n$
is 3 or 4. Let $\psi $ be the limit of a sequence $1-\xi ^{\left( \nu
\right) },$ where $\xi ^{\left( \nu \right) }$ is the sequence $\eta
^{\left( \nu \right) }$ of [\textbf{Se}], Lemma 8. This sequence has the
property that $\xi ^{\left( \nu \right) }=0$ a.e. in a neighborhood of $%
\Sigma ,$ but $\xi ^{\left( \nu \right) }$ tends to 1 a.e. and $\nabla \xi
^{\left( \nu \right) }$ tends to zero in $L^{\mu -\varepsilon }$ as $\nu $
tends to infinity. It can be shown([\textbf{O1}], (28)-(35)) that these
choices imply the inequality 
\[
\int_{B^{\prime }}\eta ^{2}\left| \nabla \left( Q^{\tau (k)/4}\right)
\right| ^{2}\ast 1\leq \int_{B^{\prime }}\left| \nabla \eta \right|
^{2}Q^{\tau (k)/2}\ast 1, 
\]
where $\tau (k)=n[n/(n-2)]^{k}.$ Taking $k$ to equal zero, the right-hand
side of this expression is bounded by the $L^{n}$-norm of $du$ over $\Omega
. $ Applying the Sobolev inequality to the left-hand side allows us to
repeat the preceding integration by parts for $k=1.$ Applying the Sobolev
inequality to the resulting inequality allows iterations for progressively
higher values of $k.$ In this way any finite $L^{p}$-norm for $du$ can be
bounded by the $L^{n}$-norm of $du$ over $\Omega .$ We conclude that $du$
lies in the space $L^{p}(B)$ for any finite value of $p$ and is an $H^{1,2}$
weak elliptic subsolution on $B^{\prime }.$ Then $u$ is H\"{o}lder
continuous and the proof is complete.

\bigskip

Let $D$ be an $n$-disc of radius $R,$ completely contained in the $n$-disc $%
B^{\prime },$ completely containing the singular set $\Sigma $ and centered
at a point $x_{0}\in \Sigma .$ We require a classical result on linear
boundary-value problems:

\begin{lemma}
If $\rho \in C^{1,\alpha }\left( D\right) $ and $w\in C^{0}\left( \partial
D\right) ,$ then $\exists v:D\rightarrow \Bbb{R}^{k}$ such that $v\in
C^{2,\alpha }\left( D\right) \cap C^{0}\left( \overline{D}\right) $ and $v$
satisfies the linear boundary-value problem 
\begin{eqnarray}
\delta \left[ \rho (\left| x\right| ^{2})dv\right] &=&0\qquad
in\,\,D_{R}(x_{0}), \\
v_{\vartheta } &=&w_{\vartheta }\qquad on\,\,\partial D,  \nonumber
\end{eqnarray}
where the subscripted $\vartheta $ denotes the tangential component of the
map in coordinates $(r,\vartheta _{1},...\vartheta _{n-1}).$
\end{lemma}

\bigskip

\textit{Proof}. Condition (15) implies [\textbf{U}] that 
\[
\rho \left( \left| x\right| ^{2}\right) \geq K 
\]
for some positive constant $K.$ This inequality implies strict ellipticity
of the linearized equations (17). The result now follows from Theorem 6.13
of [\textbf{GT}], although that result is stated for scalar equations,
because the differential operator in (17) is diagonal. This completes the
proof.

\bigskip

Define a map $\varphi :D\rightarrow \Bbb{R}^{k}$ and consider the nonlinear
boundary-value problem 
\begin{eqnarray}
\delta \left( \rho (\left| d\varphi \right| ^{2})d\varphi \right) &=&0\qquad
in\,\,D_{R}(x_{0}), \\
\varphi _{\vartheta } &=&u_{\vartheta }\qquad on\,\,\partial D.  \nonumber
\end{eqnarray}

\begin{lemma}
If $u$ satisfies the hypotheses of Theorem 2, then the boundary-value
problem (18) has a solution $\varphi $ in the space $C^{2,\beta }\left(
D\right) \cap C^{0,\alpha }\left( \overline{D}\right) .$
\end{lemma}

\textbf{Remarks.} If $u$ lies in the space $C^{2,\beta }\left( D/\Sigma
\right) $ and if $\Sigma $ is an isolated point (or by extension, a finite
point set), then the smoothness of $\varphi $ follows from Schauder
estimates, and as the radius $R$ of $D$ shrinks to a point, the boundary
conditions of problem (18) remain smooth. In this case one can compare $du$
to $d\varphi $ with the goal of applying Theorem III.1.3 of [\textbf{G}]
exactly as in [\textbf{O1}], and no further remarks are necessary. If,
however, $\Sigma $ is not a point, then for sufficiently small $R,$ $D_{r}$
will intersect $\Sigma $ and we have only the result of Lemma 3, that $u$ is
H\"{o}lder continuous on a domain that includes the singular set. It is not
explicitly shown in [\textbf{O1}] that this is sufficient boundary
regularity for completing the proof of Theorem 2; but that is in fact the
case, as we will show here.

\bigskip

\textit{Proof of Lemma 5.} Consider the boundary-value problem (17) for $w=u$
on $D_{r}(x_{0}),$ where $r\in (0,R].$ The boundary data are H\"{o}lder
continuous by Lemma 3, so the solution $v$ lies in the space $C^{2,\alpha
}\left( D_{r}\right) \cap C^{0}\left( \overline{D}_{r}\right) $ by Lemma 4.
Now we extend to systems the proof of [\textbf{LU}], Theorem 4.8.7. That is,
we solve a sequence of boundary-value problems having the form 
\begin{eqnarray}
\delta \left( \rho (\left| d\varphi \right| ^{2})d\varphi \right) &=&0\qquad
in\,\,D_{r_{i}}(x_{0}),\;D_{r_{i}}\subset D_{R}, \\
\varphi _{\vartheta } &=&v_{\vartheta }\qquad on\,\,\partial D_{r_{i}}, 
\nonumber
\end{eqnarray}
where $\left\{ r_{i}\right\} \rightarrow R.$ A $C^{2,\alpha }$ solution $%
\varphi _{i}$ to this problem exists for every $i$ by Theorem 1 of [\textbf{%
SS2}]. (The differentiability requirements on the boundary are encapsulated
in the definition of the space $\mathcal{D}_{2}$ of that paper.) Also, by
hypothesis $u$ is bounded by a constant depending only on $N.$ This gives a
uniform bound on the boundary data $v_{\vartheta }$ on each $\,\partial
D_{r_{i}}.$ Solutions $\varphi _{i}$ of (19) satisfy a maximum principle,
for each $i,$ by Sec. 2 of [\textbf{SS3}]. Thus the sequence $\left\{
\varphi _{i}\right\} $ possesses a subsequence which converges, as $r_{i}$
tends to $R,$ to a solution $\varphi \in C^{2,\beta }\left( D_{R}\right)
\cap C^{0,\alpha }\left( \overline{D}_{R}\right) ,$ as required. This
completes the proof of Lemma 5.

\bigskip

We now complete the proof of Theorem 2 by showing that the differential $du$
is H\"{o}lder continuous in a domain that includes the singular set.

For sufficiently small $B,$ we can construct a suitable $n$-disc, on the
boundary of which the tangential component of a comparison vector can be
forced to agree with the tangential component of $u$ (\textit{c.f.} [\textbf{%
Li}], Sec. 3).

Consider a solution $\varphi $ to the boundary-value problem (19). Combining
Lemma 5 with Theorem III.1.2 of [\textbf{G}], we find that if $(d\varphi
)_{R,x_{0}}$ denotes the mean value of the 1-form $d\varphi $ on $%
D_{R}(x_{0}),$ then for any sufficiently small $R,$ $d\varphi $ satisfies 
\[
\int_{D_{R}(x_{0})}\left| d\varphi -(d\varphi )_{R,x_{0}}\right| ^{2}\ast
1\leq CR^{n+2\lambda } 
\]
for some number $\lambda \in (0,1].$ Then 
\[
\int_{D_{R}(x_{0})}\left\langle d\left( u-\varphi \right) ,\left[ \rho
(\left| du\right| ^{2})du-\rho (\left| d\varphi \right| ^{2})d\varphi \right]
\right\rangle \ast 1 
\]
\[
=\int_{D_{R}(x_{0})}\left\langle \left( u-\varphi \right) ,\rho (\left|
du\right| ^{2})A(du,du)\right\rangle \ast 1. 
\]
Apply a generalized mean-value formula to the 1-form $\rho (\left| ds\right|
^{2})ds$ as in [\textbf{Si}], Lemma 1.1. We obtain 
\[
\int_{D_{R}(x_{0})}\left| d\left( u-\varphi \right) \right| ^{2}\ast 1\leq 
\]
\[
C\left( \int_{D_{R}(x_{0})}\left( \left| du\right| +\left| d\varphi \right|
\right) \left| x\right| \ast 1+\int_{D_{R}(x_{0})}\left| u-\varphi \right|
\rho (\left| du\right| ^{2})\left| u\right| \left| du\right| ^{2}\ast
1\right) 
\]
\begin{equation}
\equiv i_{1}+i_{2}.
\end{equation}
We have 
\[
i_{1}\leq \int_{D_{R}(x_{0})}\left( \left| d\left( u-\varphi \right) \right|
+2\left| d\varphi \right| \right) \left| x\right| \ast 1\leq 
\]
\[
\varepsilon \int_{D_{R}(x_{0})}\left| d\left( \varphi -u\right) \right|
^{2}\ast 1+ 
\]
\[
C\left( S^{n},\varepsilon ,\left\| d\varphi \right\| _{\infty }^{2}\right)
\int_{0}^{R}\left| x\right| ^{n+1}d\left| x\right| , 
\]
where the sup norm of $d\varphi $ depends on the modulus of continuity for $%
u $ through eq. (18). 
\[
i_{2}\leq C\left( \rho \right) \int_{D_{R}(x_{0})}\left| u-\varphi \right|
\left| u\right| \left| du\right| ^{2}\ast 1\leq 
\]
\[
R^{-\nu }\int_{D_{R}(x_{0})}\left| u-\varphi \right| ^{2}\left| u\right|
^{2}\ast 1+R^{\nu }\int_{D_{R}(x_{0})}\left| du\right| ^{4}\ast 1\leq 
\]
\[
R^{-\nu }\int_{D_{R}(x_{0})}\left| u-\varphi \right| ^{2}\left| u\right|
^{2}\ast 1+C\left( \left\| du\right\| _{4p}^{4},n\right) R^{n(p-1)/p+\nu } 
\]
for a constant $\nu $ to be chosen and $p$ so large that $\nu p>n$. \ We
have by the Sobolev Theorem 
\[
R^{-\nu }\int_{D_{R}(x_{0})}\left| u-\varphi \right| ^{2}\left| u\right|
^{2}\ast 1\leq 
\]
\[
R^{-\nu }\left( \int_{D_{R}(x_{0})}\left| u-\varphi \right| ^{2n/(n-2)}\ast
1\right) ^{(n-2)/n}\left( \int_{D_{R}(x_{0})}\left| u\right| ^{n}\ast
1\right) ^{2/n} 
\]
\[
\leq R^{-\nu }C_{Sobolev}\int_{D_{R}(x_{0})}\left| d\left( u-\varphi \right)
\right| ^{2}\ast 1\left( \int_{D_{R}(x_{0})}\left| u\right| ^{n}\ast
1\right) ^{2/n} 
\]
\[
\leq R^{-\nu }C\left\| u^{2}\right\| _{C^{0,\gamma }\left( D\right)
}\int_{D_{R}(x_{0})}\left| d\left( u-\varphi \right) \right| ^{2}\ast
1\left( \int_{0}^{R}\left| x\right| ^{n-1}d\left| x\right| \right) ^{2/n} 
\]
\[
\leq R^{2-\nu }C\int_{D_{R}(x_{0})}\left| d\left( u-\varphi \right) \right|
^{2}\ast 1. 
\]
Choose $\nu \in \left( 0,2\right) .$ Substitute the estimates for $i_{1}$
and $i_{2}$ into the right-hand side of (20) and absorb small terms on the
left in (20) to obtain 
\[
\int_{D_{R}(x_{0})}\left| d\left( u-\varphi \right) \right| ^{2}\ast 1\leq
CR^{n+\lambda ^{\prime }} 
\]
for some positive $\lambda ^{\prime }.$

The minimizing property of the mean value with respect to location
parameters implies 
\[
\int_{D_{R}(x_{0})}\left| du-(du)_{R,x_{0}}\right| ^{2}\ast 1\leq
\int_{D_{R}(x_{0})}\left| du-(d\varphi )_{R,x_{0}}\right| ^{2}\ast 1 
\]
\[
\leq \int_{D_{R}(x_{0})}\left| du-d\varphi \right| ^{2}\ast
1+\int_{D_{R}(x_{0})}\left| d\varphi -(d\varphi )_{R,x_{0}}\right| ^{2}\ast
1 
\]
\begin{equation}
\leq CR^{n+\ell }
\end{equation}
for some $\ell >0.$ Because these estimates can be repeated for any
sufficiently small value of $R,$ the proof of Theorem 2 is completed by the
local form of Campanato's Theorem (Theorem III.1.3 of [\textbf{G}]).

\subsection{Weak solutions of eqs. (11) and (15)}

If we make no assumptions about the singular set but assume that $u$
satisfies (14) weakly in $B,$ it is possible to show by estimating
difference quotients that $du$ is an element of the space $H^{1,2}(B).$ The
next logical step would be to show $du\in L^{\infty }\left( B\right) .$ This
step cannot be taken in the constrained case by following an analogy to the
unconstrained case. The latter arguments proceed from a scalar inequality,
as in Sec. 1 of [\textbf{U}]; but in order to adapt that argument to the
constrained case it is necessary to choose local coordinates on the tangent
space of $N,$ as in Theorem 2 of [\textbf{O1}]. This requires some \textit{a
priori} information about the singular set of $u.$

Theorem 2 of the preceding section implies that if an $E$-critical map $u$
is bounded and H\"{o}lder continuous on an open Euclidean domain, then $du$
is H\"{o}lder continuous on small compact subdomains. The initial continuity
assumption means that the target geometry will play little role in the
analysis beyond its contribution to the nonlinearity of the variational
equations.

\subsubsection{An alternate proof of Theorem 2}

The arguments of [\textbf{U}] imply that weak solutions of (18) are
H\"{o}lder continuous, but the H\"{o}lder estimate implied by that work
cannot be continued up to the boundary. Nevertheless, it is possible to show
that interior smoothness of weak solutions to (18) is sufficient to complete
the proof of Theorem 2 using a modification of the preceding arguments.

The weak form of problem (18) can be written

\begin{equation}
\int_{D_{R}(x_{0})}\left\langle d\zeta ,\rho \left( \left| d\varphi \right|
^{2}\right) d\varphi \right\rangle \ast 1=0,
\end{equation}
where $d\zeta $ is a closed 1-form in $L^{2}(D_{R})$ having vanishing
tangential component on $\partial D_{R}.$ Applying standard
function-theoretic arguments on $\Bbb{R}^{n},$ we consider $\zeta $ to be an
admissible test function; \textit{c.f.} eq. (1.2) of [\textbf{Si}]. Writing
(22) as the weak variational equations of the energy functional $E$ with $N$
replaced by $\Bbb{R}^{k},$ we have 
\[
\int_{D_{R}(x_{0})}\int_{0}^{\left| d\varphi \right| ^{2}}\rho \left(
s\right) ds\ast 1\geq K\int_{D_{R}(x_{0})}\left| d\varphi \right| ^{2}\ast
1, 
\]
so $d\varphi $ lies in the space $L^{2}(B_{R})$ by ellipticity and finite
energy. (See also Sec. 1 of [\textbf{U}].) The proof of Lemma 3 implies that 
$du$ lies in $L^{2}(D_{R}).$ Because $d\left( u-\varphi \right) $ is in $%
L^{2},$ we can choose $\zeta =u-\varphi $ in (22). The resulting weak
Dirichlet problem is solvable by Proposition 4.3 of [\textbf{Si}]; see also [%
\textbf{ISS}]. The 1-form $d\varphi $ is H\"{o}lder continuous in the
interior of $D$ by Proposition 4.4 of [\textbf{Si}], which is derived from [%
\textbf{U}]. \ The Campanato Theorem implies that 
\[
\int_{D_{R/2}(x_{0})}\left| d\varphi -(d\varphi )_{R/2,x_{0}}\right|
^{2}\ast 1\leq CR^{n+\alpha } 
\]
for some $\alpha \in (0,2].$ Estimating (20) as in the preceding section, we
find that 
\[
\int_{D_{R}(x_{0})}\left| d\left( u-\varphi \right) \right| ^{2}\ast 1\leq
CR^{n+\mu } 
\]
for some positive $\mu .$ Then of course 
\[
\int_{D_{R/2}(x_{0})}\left| d\left( u-\varphi \right) \right| ^{2}\ast 1\leq
CR^{n+\mu }. 
\]
Rewrite inequality (21) over $D_{R/2}(x_{0})$ to obtain 
\[
\int_{D_{R/2}(x_{0})}\left| du-(du)_{R/2,x_{0}}\right| ^{2}\ast 1\leq
\int_{D_{R/2}(x_{0})}\left| du-(d\varphi )_{R/2,x_{0}}\right| ^{2}\ast 1 
\]
\[
\leq \int_{D_{R/2}(x_{0})}\left| du-d\varphi \right| ^{2}\ast
1+\int_{D_{R/2}(x_{0})}\left| d\varphi -(d\varphi )_{R/2,x_{0}}\right|
^{2}\ast 1 
\]
\[
\leq CR^{n+\ell } 
\]
for some $\ell >0.$ This completes the alternate proof of Theorem 2.

An application of this argument to an unconstrained problem for bundle
curvature is given in [\textbf{O2}].

\section{Nonuniformly elliptic solutions having nonzero vorticity}

Note that the map $u$ enters into the problem of the preceding section only
through its geometry. The variational equations in the unconstrained case
are written in terms of $du,$ and the map $u$ does not directly appear in
them. This raises the question of how much of the theory can be deduced in
the unconstrained case without assuming the existence of a potential.

Thus we consider systems having the form [\textbf{O1}] 
\begin{equation}
\delta \left( \rho \left( Q\right) \omega \right) =0,
\end{equation}
\begin{equation}
d\omega =\nu \wedge \omega ,
\end{equation}
for $\nu $ $\in \Lambda ^{1}(V),$ where $V$ is a smooth section of a vector
bundle over an open, bounded domain $\Omega $ of $\Bbb{R}^{n};$ $\omega \in
\Lambda ^{p}\left( V\right) ;$ $Q=\omega \wedge \ast \omega ,$ where $\ast
:\Lambda ^{p}\rightarrow \Lambda ^{n-p}$ is the Hodge involution; $\rho $ is
defined as in the preceding sections, but will be assumed to satisfy an
inequality somewhat different from (1).

The condition 
\begin{equation}
d\omega =0
\end{equation}
implies, by the converse of the Poincar\'{e} Lemma, the local existence of a
potential $u\in \Lambda ^{p-1}\left( V\right) $ such that $du=\omega .$ \
Thus solutions of (25) lie in a cohomology class, whereas solutions of (24)
do not in general. However, the integrability condition (24) generates a
closed ideal when $p=1.$ \ \ Obvious modifications of condition (24)
generate a closed ideal for solutions of higher degree (see, \textit{e.g.}, [%
\textbf{Ed}], Theorem 4-2.1).

\ In the interpretation of eqs. (23), (25) in which $\omega $ is dual to the
flow velocity, the vanishing of the \textit{vorticity} $d\omega $ expresses
the property that $\omega $ must integrate to zero along any curve
homologous to zero. The 0-form $u$ is the flow potential. One-form solutions
to (24) have only the weaker property that $\omega =\ell du$ for some 0-form 
$\ell .$

\bigskip

\textbf{Remark on terminology}. The distinction between \textit{curl-free}
and \textit{rotation-free} fields is sometimes used to characterize velocity
fields corresponding to (25) and (24), respectively (\textit{c.f.} pp. 123,
124 of [\textbf{MTW}]; p. 28 of [\textbf{So}]). In [\textbf{O1}] the term 
\textit{irrotational field} is used to denote a curl-free field, and any
field which is not curl-free is called \textit{rotational.} While that
terminology may be misleading physically, the term \textit{curl-free} is not
mathematically correct in higher dimensions, so either choice of terms is
open to criticism.

\bigskip

In Theorem 7 and Corollary 8 of [\textbf{O1}] a H\"{o}lder estimate is
derived for the variant (23), (24) of the nonlinear Hodge equations on a
possibly singular domain. \ As the solution approaches the critical value at
which the ellipticity of the differential operator breaks down, the elliptic
estimate of [\textbf{O1}] also breaks down. \ In this section we derive an
estimate which is uniform over the entire subcritical range.

We assume that $\omega $ is a classical solution of eqs. (23), (24) outside
a singular set of prescribed dimension and that the density $\rho $
satisfies 
\begin{equation}
\kappa _{5}^{-1}(Q+k_{0})^{q}\leq \rho (Q)+2Q\rho ^{\prime }(Q)\;\leq \kappa
_{5}(Q+k_{0})^{q},
\end{equation}
for constants $\kappa _{5},q>0\;$and $k_{0}\geq 0.$ \ Condition (26) was
introduced in [\textbf{U}] in connection with a generalized version of eqs.
(23), (25).\ That condition implies that there is a possibly larger value of 
$\kappa _{5}$ for which 
\begin{equation}
\kappa _{5}^{-1}(Q+k_{0})^{q}\leq \rho (Q)\;\leq \kappa _{5}(Q+k_{0})^{q}
\end{equation}
and 
\begin{equation}
\left| Q\rho ^{\prime }(Q)\right| \leq \kappa _{5}(Q+k)^{q}.
\end{equation}
In the sequel we denote by $\kappa $ a number so large that it satisfies
(26), (27), and (28). \ Condition (26) is an ellipticity condition for eqs.
(23). \ If $k=0$, then ellipticity degenerates as $Q$ tends to zero;
condition (27) implies that the density $\rho $ also tends to zero
(cavitates) in this limit. \ Thus ellipticity and noncavitation are
equivalent under condition (26). \ In applications to compressible flow, the
degeneration of ellipticity need not imply cavitation, and in cases in which
these two phenomena are equivalent, as in the Chaplygin approximation, the
degeneracy occurs at infinity rather than at zero. \ Moreover, condition
(26) is not associated with a sonic transition. \ For these reasons,
condition (26) does not appear to be appropriate for applications to fluid
dynamics. \ However, it arises in certain natural generalizations of the
Dirichlet energy; see [\textbf{HL}] and the references cited therein for
details.

The methods used to study eq. (23) also apply to systems in which (23) is
replaced by an equation of the form 
\begin{equation}
\delta \left( \rho \left( Q\right) \omega \right) =\xi \left( \omega \right)
,
\end{equation}
where $p=1$ and $\xi $ is a scalar function of $\omega $ satisfying 
\begin{equation}
\left| \xi ^{\prime }\left( \omega \right) \right| \leq \kappa \left(
Q+k\right) ^{\alpha }
\end{equation}
for $\alpha \in \Bbb{R}^{\Bbb{+}}.$ \ For simplicity we take $\alpha =q.$ \
Obvious algebraic modifications lead to results analogous to inequality
(34), Theorem 6, and Corollary 7 for general $\alpha >0.$ \ In that case,
inequality (34) may no longer be linear in its terms of zero order.

Certain properties of eqs. (29), (24) can be obtained by deriving a
differential inequality for an appropriate scalar function of the solution.
\ The case $\nu =\xi =0$ is due to Uhlenbeck [\textbf{U}], who framed the
argument in the context of a broadly defined elliptic complex. \ We
initially present a version of Uhlenbeck's argument in simpler notation for
solutions of (29) and (25), and then indicate how to modify the proof for
the case of solutions of the system (29), (24).

Denote by $H(Q)$ a $C^{1}$ function of $Q$ such that 
\[
H^{\prime }(Q)=\frac{1}{2}\rho (Q)+Q\rho ^{\prime }(Q). 
\]
Then [\textbf{U}] 
\[
\left\langle \omega ,\Delta \left( \rho (Q)\omega \right) \right\rangle
=\partial _{i}\left\langle \omega ,\partial _{i}\left( \rho (Q)\omega
\right) \right\rangle -\left\langle \partial _{i}\omega ,\partial _{i}\left(
\rho (Q)\omega \right) \right\rangle 
\]
\[
=\Delta H(Q)-\left[ \rho (Q)\left\langle \partial _{i}\omega ,\partial
_{i}\omega \right\rangle +\rho ^{\prime }(Q)\left\langle \partial _{i}\omega
,\omega \right\rangle \partial _{i}Q\right] , 
\]
where $\partial _{i}=\partial /\partial x^{i},x=x^{1},...,x^{n}\in \Omega $,
and 
\[
\rho ^{\prime }(Q)\left\langle \partial _{i}\omega ,\omega \right\rangle
\partial _{i}Q=\sum_{i}2\rho ^{\prime }(Q)\left\langle \partial _{i}\omega
,\omega \right\rangle ^{2}. 
\]
We have [for either sign of $\rho ^{\prime }\left( Q\right) $] 
\[
\rho (Q)\left\langle \partial _{i}\omega ,\partial _{i}\omega \right\rangle
+\rho ^{\prime }(Q)\left\langle \partial _{i}\omega ,\omega \right\rangle
\partial _{i}Q\geq \kappa ^{-1}(Q+k)^{q}\left| \nabla \omega \right| ^{2} 
\]
and 
\[
\left\langle \omega ,\Delta \left( \rho (Q)\omega \right) \right\rangle \leq
\Delta H(Q)-\kappa ^{-1}(Q+k)^{q}\left| \nabla \omega \right| ^{2}. 
\]
In addition, 
\[
\left\langle \omega ,\Delta \left( \rho (Q)\omega \right) \right\rangle
=\left\langle \omega ,\delta d\left( \rho \left( Q\right) \omega \right)
\right\rangle +\left\langle \omega ,d\delta \left( \rho \left( Q\right)
\omega \right) \right\rangle 
\]
\[
=\left\langle \omega ,\delta d\left( \rho \left( Q\right) \omega \right)
\right\rangle +\left\langle \omega ,d\xi \right\rangle 
\]
for solutions of (29) and (25), yielding 
\begin{equation}
\kappa ^{-1}(Q+k)^{q}\left| \nabla \omega \right| ^{2}\leq \Delta
H(Q)-\left\langle \omega ,\delta d\left( \rho \left( Q\right) \omega \right)
\right\rangle -\left\langle \omega ,d\xi \right\rangle .
\end{equation}

Define a map $\beta _{\omega }:\Lambda ^{0}\rightarrow \Lambda ^{p+1}$ by
the formula $\beta _{\omega }:\zeta \rightarrow d\zeta \wedge \omega ,$ for $%
\zeta \in \Lambda ^{0}$ and $\omega \in \Lambda ^{p}.$ \ If $\nu =0,$%
\[
\beta _{\omega }\left( \zeta \right) =d\left( \zeta \omega \right) , 
\]
but we do not use this property. \ Define the map $\beta _{\omega }^{\ast
}:\Lambda ^{p+1}\rightarrow \Lambda ^{0}$ by the formula 
\[
\beta _{\omega }^{\ast }\left( \varsigma \right) =\delta \ast \left( \omega
\wedge \ast \varsigma \right) 
\]
for $\varsigma \in \Lambda ^{p+1}$. \ Writing 
\[
\left\langle \omega ,\delta d\left( \rho \left( Q\right) \omega \right)
\right\rangle \equiv \ast \left[ \omega \wedge \ast \delta d\left( \rho
\left( Q\right) \omega \right) \right] 
\]
\[
=\ast d\left[ \omega \wedge \ast \left( \rho ^{\prime }\left( Q\right)
dQ\wedge \omega \right) \right] =\delta \ast \left[ \omega \wedge \ast
\left( \rho ^{\prime }\left( Q\right) dQ\wedge \omega \right) \right] 
\]
(\textit{c.f.} Lemma 2.1.4 of [\textbf{J}]), we can write (31) in the form 
\[
\kappa ^{-1}(Q+k)^{q}\left| \nabla \omega \right| ^{2}\leq \Delta H\left(
Q\right) -\beta _{\omega }^{\ast }\beta _{\omega }\left[ \rho \left(
Q\right) \right] -\left\langle \omega ,d\xi \right\rangle . 
\]
Because 
\[
dQ=\frac{dH}{H^{\prime }\left( Q\right) }, 
\]
we can rewrite this inequality, in terms of $H$, as 
\[
\kappa ^{-1}(Q+k)^{q}\left| \nabla \omega \right| ^{2}\leq \Delta
H-div\left\{ \ast \left[ \omega \wedge \ast \left( \frac{\rho ^{\prime
}\left( Q\right) }{H^{\prime }\left( Q\right) }dH\wedge \omega \right) %
\right] \right\} 
\]
\[
-\left\langle \omega ,d\xi \right\rangle 
\]
\[
=\Delta H-\beta _{\omega }^{\ast }\left[ \varsigma _{\omega }\left( H\right) %
\right] -\left\langle \omega ,d\xi \right\rangle . 
\]
for 
\[
\varsigma _{\omega }\left( H\right) =\frac{\rho ^{\prime }\left( Q\right) }{%
H^{\prime }\left( Q\right) }dH\wedge \omega . 
\]
Write 
\[
L_{\omega }\left( H\right) \equiv \Delta H-\beta _{\omega }^{\ast }\left[
\varsigma _{\omega }\left( H\right) \right] 
\]
\[
=\sum_{k,j}\partial _{k}\left( a_{kj}\partial _{j}\right) H. 
\]
If $\rho ^{\prime }\left( Q\right) $ is nonpositive, then the matrix $a_{kj}$
satisfies 
\[
1\leq a_{kj}=1+\frac{Q\left| \rho ^{\prime }\left( Q\right) \right| }{%
H^{\prime }\left( Q\right) }\leq 1+\frac{2\kappa (Q+k)^{q}}{\kappa
^{-1}(Q+k)^{q}}. 
\]
Letting $\pi =\left( \pi _{1},\ldots ,\pi _{n}\right) $ denote an $n$%
-vector, we have 
\[
\left| \pi \right| ^{2}\leq \sum_{k,j}\pi _{k}a_{kj}\pi _{j}\leq \left(
1+2\kappa ^{2}\right) \left| \pi \right| ^{2}. 
\]
If $\rho ^{\prime }\left( Q\right) >0,$ write 
\[
div\left[ \left( 1-\frac{Q\rho ^{\prime }\left( Q\right) }{H^{\prime }\left(
Q\right) }\right) grad\left( H\right) \right] = 
\]
\[
div\left[ \left( \frac{\frac{1}{2}\rho \left( Q\right) +Q\rho ^{\prime
}\left( Q\right) -Q\rho ^{\prime }\left( Q\right) }{H^{\prime }\left(
Q\right) }\right) grad\left( H\right) \right] 
\]
\[
=div\left[ \left( \frac{\rho \left( Q\right) }{2H^{\prime }\left( Q\right) }%
\right) grad\left( H\right) \right] . 
\]
The matrix $a_{kj}$ now satisfies 
\[
\frac{\kappa ^{-1}}{2\kappa }\leq \frac{\rho \left( Q\right) }{2H^{\prime
}\left( Q\right) }=a_{kj}\leq \frac{2H^{\prime }\left( Q\right) }{2H^{\prime
}\left( Q\right) }=1. 
\]
Letting $\pi =\left( \pi _{1},\ldots ,\pi _{n}\right) $ denote an $n$%
-vector, we have 
\[
\frac{\left| \pi \right| ^{2}}{2\kappa ^{2}}\leq \sum_{k,j}\pi _{k}a_{kj}\pi
_{j}\leq \left| \pi \right| ^{2}. 
\]
Thus $L$ is a uniformly elliptic operator on $H$ for either sign of $\rho
^{\prime }\left( Q\right) .$

It remains to estimate the lower-order nonlinear term $\left\langle \omega
,d\xi \right\rangle $ and to adjust for $\nu \neq 0.$ \ We have 
\[
\left\langle \omega ,d\xi \right\rangle =\left\langle \omega ,\xi ^{\prime
}\left( \omega \right) d\omega \right\rangle =\xi ^{\prime }\left( \omega
\right) \left\langle \omega ,\nu \wedge \omega \right\rangle \geq 
\]
\[
-\left| \xi ^{\prime }\left( \omega \right) \right| \left| \nu \right| Q\geq
-\left| \xi ^{\prime }\left( \omega \right) \right| \left| \nu \right|
\left( Q+k\right) \geq -\kappa \left| \nu \right| \left( Q+k\right) ^{q+1}, 
\]
using (30) with $\alpha =q.$ Integrating condition (26) over a dummy
variable in $\left[ 0,Q\right] $ and using $H(0)=0$, we find that 
\begin{equation}
\kappa ^{-1}(Q+k)^{q+1}\leq H(Q)\;\leq \kappa (Q+k)^{q+1}
\end{equation}
and obtain 
\[
\left\langle \omega ,d\xi \right\rangle \geq -\kappa ^{2}\left| \nu \right|
H(Q). 
\]

In the proof of Theorem 7 of [\textbf{O1}] it was shown that if $\omega ,\nu 
$ smoothly satisfies (23), (24), then there is an independent positive
constant $C$ and a sufficiently small constant $\varepsilon \left( \kappa
\right) $ for which 
\[
0\leq (\kappa ^{-1}-\varepsilon \kappa )(Q+k)^{q}|\nabla \omega |^{2} 
\]
\[
\leq \Delta H(Q)-\ast d\left[ \omega \wedge \ast \left( \rho ^{\prime
}(Q)dQ\wedge \omega \right) \right] 
\]
\begin{equation}
+C(Q+k)^{q}\left( |\nabla v|+|v|^{2}\right) Q.
\end{equation}
\ We can convert this estimate to an inequality in $H$, noticing first that 
\[
(Q+k)^{q}\left( |\nabla v|+|v|^{2}\right) Q\leq \kappa \left( |\nabla
v|+|v|^{2}\right) H\left( Q\right) 
\]
by (32). \ Now taking into account the term $-\left\langle \omega ,d\xi
\right\rangle $ and reasoning as in\ the curl-free case, we rewrite (33) in
the form 
\begin{equation}
0\leq L_{\omega }\left( H\right) +C\left( \kappa ,q\right) \left( |\nabla
v|+|v|^{2}+t|v|\right) H,
\end{equation}
where 
\[
L_{\omega }\left( H\right) =\Delta H-div\left\{ \ast \left[ \omega \wedge
\ast \left( \frac{\rho ^{\prime }\left( Q\right) }{H^{\prime }\left(
Q\right) }dH\wedge \omega \right) \right] \right\} 
\]
and $t=0$ unless $\xi $ is nonzero, in which case $t=1.$ \ This operator is
clearly elliptic on $H$, as we did not use the closure of $\omega $ under $d$
in establishing uniform ellipticity for the corresponding operator in the
case $\nu =0$.

Notice that the operator $L_{\omega }\left( H\right) $ can be written as an
operator on $Q$ having the form 
\[
\widetilde{L}_{\omega }\left( Q\right) =\partial _{i}\left[ \left( \frac{1}{2%
}\rho (Q)+Q\rho ^{\prime }(Q)\right) \partial _{i}Q\right] 
\]
\[
-\ast d\left[ \omega \wedge \ast \left( \rho ^{\prime }(Q)dQ\wedge \omega
\right) \right] . 
\]
This operator is only elliptic on $Q$ only if $k$ exceeds zero. \ Thus for
example, inequality (34) allows us to extend Corollary 8 of [\textbf{O1}],
which was based on an elliptic inequality for $\widetilde{L}_{\omega }\left(
Q\right) $. \ The bound on $\omega $ established in that result is not
uniform as $Q$ tends to zero unless the constant $k$ in condition (26)
exceeds zero. \ We can remove that restriction if we place different $L^{p}$
hypotheses on the solution. \ In comparison with the hypotheses of [\textbf{%
O1}], Corollary 8, the new $L^{p}$ hypotheses placed directly on $\omega $
are somewhat stronger, whereas those placed indirectly on $\omega ,$ through
the hypothesis on $v$ and its derivatives, are considerably weaker.

\begin{theorem}
Let the pair $\omega ,\nu $ smoothly satisfy eqs. (23), (24), with $\rho $
satsifying condition (26), on a domain $\Omega /\Sigma .$ Let $\Omega $ be a
type-A domain of $\Bbb{R}^{n},$ $n>2.$ \ Let $\Sigma $ be a compact singular
set, completely contained in a sufficiently small $n$-disc $B$, which is
itself completely contained in $\Omega .$ If $n$ exceeds 4, let $2n/\left(
n-2\right) <\mu <n,$ where $\mu $ is the codimension of $\Sigma ,$ and let $%
\omega $ lie in $L^{n\left( q+1\right) }\left( B\right) .$ \ If $n$ = 3 or
4, let $\omega $ lie in $L^{4\left( q+1\right) \beta \gamma _{1}}\left(
B\right) \cap L^{4\left( q+1\right) \gamma _{2}}\left( B\right) ,$ where $%
\beta =\left( \mu -\varepsilon \right) /\left( \mu -2-\varepsilon \right) $
for $2<\mu \leq n,$ $\varepsilon >0,$ and $\frac{1}{2}<\gamma _{1}<\gamma
_{2}.$ If $|\nabla v|+|v|^{2}\in L^{s}\left( B\right) $ for some $s$
exceeding $n/2$, then $\omega $ is bounded on compact subdomains of $\Omega
. $
\end{theorem}

\textit{Proof}. \ Integrate inequality (34) against the Serrin test function
as in Lemma 3 of the preceding section. Using (32), the $L^{p}$ hypothesis
on $\omega $ translates into $L^{p}$ hypotheses on $H$ which are sufficient
for applying the arguments of Lemma 3 to $H.$ These yield an integral
inequality which can be iterated. \ After a finite number of iterations, we
find that $H$ is in $L^{P}$ for all finite values of $P$ and is a weak $%
H^{1,2}$ subsolution on $B\cap \Sigma .$ \ Theorem 5.3.1 of [\textbf{Mo]}
implies that $H$ is bounded on compact subdomains of $\Omega .$ \ Condition
(26) extends this result to $Q$, and thus to $\omega .$ \ This completes the
proof of Theorem 6.

\begin{corollary}
Assume the conditions of Theorem 6, except let $\omega $ be a 1-form,
replace eq. (23) with eq. (29), and let $\xi $ satisfy (30). \ Then the
conclusion of Theorem 6 remains valid.
\end{corollary}

\textit{Proof}. \ Clearly, $|\nabla v|+|v|^{2}+|v|\in L^{s}\left( B\right) $
for some $s$ exceeding $n/2.$ \ This completes the proof of Corollary 7.

\bigskip

\textbf{Remark}. We take this opportunity to correct a pair of misprints in
the statement of Corollary 8 of [\textbf{O1}]: replace $L^{4q}(B)$ by $%
L^{4q_{1}}(B)$ and $1/2<q_{0}<q$ by $1/2<q_{0}<q_{1}.$

\bigskip

\begin{center}
{\large Literature cited}
\end{center}

[\textbf{A}] M. Ara, Geometry of F-harmonic maps, \textit{Kodai Math. J.} 
\textbf{22} (1999), 243-263.

[\textbf{Ba}] H. Bateman, Notes on a differential equation which occurs in
the two-dimensional motion of a compressible fluid and the associated
variational problem, \textit{Proc. R. Soc. London Ser. A,} \textbf{125}
(1929), 598--618.

\textbf{[Be] \ }L. Bers, \textit{Mathematical Aspects of Subsonic and
Transonic Gas Dynamics,} Wiley, New York, 1958.

[\textbf{Ch}] C. J. Chapman, \textit{High Speed Flow}, Cambridge University
Press, Cambridge, 2000.

[\textbf{CF}] G-Q. Chen and M. Feldman, Multidimensional transonic shocks
and free boundary problems for nonlinear equations of mixed type, preprint.

[\textbf{CL}] D. Costa and G. Liao, On removability of a singular
submanifold for weakly harmonic maps, \textit{J. Fac. Sci. Univ. Tokyo Sect.
1A Math.} \textbf{35} (1988), 321-344.

[\textbf{D}] \ E. DiBenedetto, $C^{1+\alpha }$ local regularity of weak
solutions of degenerate elliptic equations, \textit{Nonlinear Analysis T. M.
A.} \textbf{7}, No. 8 (1983), 827-850.

[\textbf{DO}] G. Dong and B. Ou, Subsonic flows around a body in space, 
\textit{Commun. Partial Differential Equations} \textbf{18} (1993), 355-379.

\textbf{[Ed]} D. G. B. Edelen, \textit{Applied Exterior Calculus}, Wiley,
New York, 1985.

[\textbf{EL}] J. Eells and L. Lemaire, Some properties of exponentially
harmonic maps, \textit{Proc. Banach Center, Semester on PDE} \textbf{27}
(1990), 127-136.

[\textbf{EP}] J. Eels and J. C. Polking, Removable singularities of harmonic
maps, \textit{Indiana Univ. Math. J.} \textbf{33}, No. 6 (1984), 859-871.

[\textbf{Ev}] L. C. Evans, A new proof of local $C^{1+\alpha }$ regularity
for solutions of certain degenerate elliptic P.D.E., \textit{J. Differential
Equations} \textbf{45} (1982), 356-373.

[\textbf{F}] M. Fuchs, \textit{Topics in the Calculus of Variations},
Vieweg, Wiesbaden, 1994.

[\textbf{FH}] \ N. Fusco and J. Hutchinson, Partial regularity for
minimisers of certain functionals having nonquadratic growth, \textit{Ann.
Mat. Pura Appl.} \textbf{155} (1989), 1-24.

[\textbf{G}] M. Giaquinta, \textit{Multiple Integrals in the Calculus of
Variations and Nonlinear Elliptic Theory}, Princeton University Press,
Princeton, 1983.

[\textbf{GT}] D. Gilbarg and N. S. Trudinger, \textit{Elliptic Partial
Differential Equations of Second Order,} Springer-Verlag, Berlin, 1983.

[\textbf{HL}] \ R. Hardt and F-H. Lin, Mappings minimizing the $L^{p}$ norm
of the gradient, \textit{Commun. Pure Appl. Math.} \textbf{40 }(1987)\textbf{%
,} 555-588.

[\textbf{HJW}] S. Hildebrandt, J. Jost, and K.-O. Widman, Harmonic mappings
and minimal surfaces, \textit{Inventiones Math.} \textbf{62} (1980), 269-298.

[\textbf{ISS}] T. Iwaniec, C. Scott, and B. Stroffolini, Nonlinear Hodge
theory on manifolds with boundary, \textit{Annali Mat. Pura Appl.} (4) 
\textbf{177} (1999), 37-115.

[\textbf{J}] J. Jost, \textit{Riemannian Geometry and Geometric Analysis},
Springer-Verlag, Berlin, 1995.

[\textbf{KFL}] A. D. Kanfon, A. F\"{u}zfa, and D. Lambert, Some examples of
exponentially harmonic maps, arXiv:math-ph/0205021.

[\textbf{LU}] \ O. A. Ladyzhenskaya and N. N. Ural'tseva, \textit{Linear and
Quasilinear Elliptic Equations,} Academic Press, New York, 1968.

[\textbf{Le}] \ J. L. Lewis, Regularity of the derivatives of solutions to
certain degenerate elliptic equations, \textit{Indiana Univ. Math J. 32
(1983), 849-858.}

[\textbf{Li}] G. Liao, A regularity theorem for harmonic maps with small
energy, \textit{J. Differential Geometry} \textbf{22} (1985), 233-241.

[\textbf{LM}] E. Loubeau and S. Montaldo, A note on exponentially harmonic
morphisms, \textit{Glasgow Math. J.} \textbf{42} (2000), 25-29.

[\textbf{Me}] M. Meier, Removable singularities of harmonic maps and an
application to minimal submanifolds, \textit{Indiana Univ. Math. J.} \textbf{%
35}, No. 4 (1986), 705-726.

[\textbf{MTW}] C. W. Misner, K. S. Thorne, and J. A. Wheeler, \textit{%
Gravitation,} Freeman, New York, 1973.

[\textbf{Mo}] \ C. B. Morrey, Jr., \textit{Multiple Integrals in the
Calculus of Variations}, Springer-Verlag, Berlin, 1966.

[\textbf{O1}] T. H. Otway, Nonlinear Hodge maps, \textit{J. Math. Phys.} 
\textbf{41}, No. 8 (2000), 5745-5766.

[\textbf{O2}] T. H. Otway, Uniformly and nonuniformly elliptic variational
equations with gauge invariance, arXiv:math-ph/0007028.

[\textbf{SaU}] J. Sacks and K. Uhlenbeck, The existence of minimal
immersions of 2-spheres, \textit{Ann. of Math.} (2) \textbf{113} (1981),
1-24.

[\textbf{Sch}] R. Schoen, Analytic aspects of the harmonic map problem, in:
S. S. Chern, ed., \textit{Seminar on Nonlinear Partial Differential Equations%
}, Springer-Verlag, New York, 1985, pp. 321-358.

[\textbf{ScU}] R. Schoen and K. Uhlenbeck, A regularity theory for harmonic
maps, \textit{J. Diff. Geom.} \textbf{17} (1982), 307-335.

[\textbf{Sed}] V. I. Sedov, \textit{Introduction to the Mechanics of a
Continuous Medium}, Addison-Wesley, Reading, 1965.

[\textbf{Se}] J. Serrin, Local behavior of solutions of quasilinear
equations, \textit{Acta Math.} \textbf{111} (1964), 247-302.

[\textbf{Sh}] \ M. Shiffman, On the existence of subsonic flows of a
compressible fluid, \textit{J. Rat. Mech. Anal.} \textbf{1} (1952), 605-652.

[\textbf{Si}] L. M. Sibner, An existence theorem for a nonregular
variational problem, \textit{Manuscripta} \textit{Math.} \textbf{43}, 45-72%
\textbf{\ }(1983).

[\textbf{SS1}] L. M. Sibner and R. J. Sibner, A nonlinear Hodge-de Rham
theorem, \textit{Acta Math.} \textbf{125} (1970), 57-73.

\textbf{[SS2]} \ L. M. Sibner and R. J. Sibner, Nonlinear Hodge theory:
Applications, \textit{Advances in Math.} \textbf{31} (1979), 1-15.

\textbf{[SS3]} \ L. M. Sibner and R. J. Sibner, A sub-elliptic estimate for
a class of invariantly defined elliptic systems, \textit{Pacific J. Math.} 
\textbf{94}, No. 2 (1982), 417-421.

[\textbf{Sm}] \ P. D. Smith, Nonlinear Hodge theory on punctured Riemannian
manifolds, \textit{Indiana Univ. Math. J.} \textbf{31}, No. 4 (1982),
553-577.

[\textbf{So}] C. F. Sopuerta, Applications of timelike and null congruences
to the construction of cosmological models, Ph.D. Thesis, Universitat de
Barcelona, 1996.

[\textbf{T}] G. E. Tanyi, On the critical points of the classical elastic
energy functional, \textit{Afrika Matematika} \textbf{1 }(1978), 35-43.

\textbf{[U]} K. K. Uhlenbeck, Regularity for a class of nonlinear elliptic
systems, \textit{Acta Math.} \textbf{138 }(1977), 219-240.

\end{document}